\documentclass[aps,prl,twocolumn,superscriptaddress,nofootinbib]{revtex4-2}
\usepackage{amsmath}
\usepackage{orcidlink}
\usepackage{xcolor}
\hypersetup{
    colorlinks=true,
    linkcolor=blue,
    filecolor=blue,
    urlcolor=blue,
    citecolor=blue
}

\begin{document}

\title{Mechanism of Ionization Avalanche in Tokamak Microwave Gas Breakdown}

\author{Jinwoo Gwak\orcidlink{0000-0003-4767-3328}}
\affiliation{Department of Nuclear Engineering, Seoul National University, Seoul, Republic of Korea}

\author{Yeongsun Lee\orcidlink{0000-0003-4474-416X}}
\affiliation{Department of Nuclear Engineering, Seoul National University, Seoul, Republic of Korea}
\affiliation{Nuclear Research Institute for Future Technology and Policy, Seoul National University, Seoul, Republic of Korea}

\author{Jeongwon Lee\orcidlink{0000-0002-2353-2603}}
\affiliation{Korea Institute of Fusion Energy, Daejeon, Republic of Korea}

\author{Won Ik Jeong\orcidlink{0000-0001-8475-2610}}
\affiliation{Korea Institute of Fusion Energy, Daejeon, Republic of Korea}

\author{Hyun-Tae Kim\orcidlink{0009-0008-2549-5624}}
\affiliation{United Kingdom Atomic Energy Authority, Culham Campus, Abingdon, Oxfordshire OX14 3DB, United Kingdom}

\author{Yong-Seok Hwang}
\affiliation{Department of Nuclear Engineering, Seoul National University, Seoul, Republic of Korea}

\author{Min-Gu Yoo\orcidlink{0000-0002-9244-7066}}
\email[Contact author: ]{yoom@fusion.gat.com}
\affiliation{General Atomics, San Diego, CA 92186-5608, USA}

\author{Yong-Su Na\orcidlink{0000-0001-7270-3846}}
\email[Contact author: ]{ysna@snu.ac.kr}
\affiliation{Department of Nuclear Engineering, Seoul National University, Seoul, Republic of Korea}

\date{September 17, 2026}

\begin{abstract}
Microwave breakdown driven by electron cyclotron (EC) waves provides a non-inductive route to plasma initiation in reactor-scale tokamaks. We introduce a three-dimensional Monte Carlo simulation that, for the first time, self-consistently treats nonlinear wave--particle interactions, atomic collisions, and guiding-center transport. The Monte Carlo simulation unveils the key role of parallel Brownian motion in the ionization avalanche mechanism. The predicted breakdown boundary is validated against KSTAR experiments. This work concludes that microwave gas breakdown will be successful under ITER-relevant conditions at a \(\mathrm{D}_2\) prefill pressure near $2\,\mathrm{mPa}$ with $1\,\mathrm{MW}$ of injected EC power.
\end{abstract}

\maketitle

\textit{Introduction}.---Tokamak startup conventionally relies on a toroidal electric field induced by the central solenoid to achieve gas breakdown and drive the plasma current $I_p$. In reactor-scale superconducting tokamaks, however, the inductive field can generate runaway electrons, potentially leading to complete startup failure \cite{Knoepfel1975,YeongsunLee2024}. Recent studies have raised another concern that reaching high $I_p$ in compact tokamaks would be challenging due to a limited flux swing and intrinsic magnetohydrodynamic stability constraints \cite{Fitzpatrick2026,Boozer2026}. As a promising alternative, electron cyclotron (EC) waves can resolve these concerns by initiating a plasma even before the inductive field is applied \cite{kubo1983,toi1984}. This reduces inductive flux consumption during startup \cite{Forest1992,Hwang1996,Shiraiwa2004,Yoshinaga2006,Sauter2000} and prevents the inductive creation of runaway electrons. The gas breakdown phase, referred to as microwave gas breakdown hereafter,\footnote{We use the conventional term ``microwave'' although the relevant frequencies extend into the millimeter-wave range.} has been experimentally investigated to support extrapolation of breakdown requirements to future devices \cite{Jackson2007,Bae2008,Stober2011,Yoneda2017,JeongwonLee2017}. Ref.~\cite{Jackson2007} proposed a phenomenological picture in which the ionization source and particle loss rates could be separated, analogous to Townsend theory, but the underlying physical mechanisms governing these rates are still unclear. Quantifying these rates has therefore remained a critical open problem although it is essential for establishing confidence in the feasibility of microwave breakdown in future tokamaks.

The mechanism by which cold seed electrons gain sufficient energy to trigger ionizing events was first pointed out in Ref.~\cite{Suvorov1988} and recently applied to microwave gas breakdown in Ref.~\cite{Farina2018}. In a stellarator with closed flux surfaces, a subsequent ionizing avalanche can be accounted for by a macroscopic fluid description when rapid parallel particle motions facilitate equilibration and plasma parameters are accordingly well-defined \cite{Johansson2026}. In a tokamak with open field lines, however, parallel stochastic motions drive non-equilibrating diffusion along the field lines, necessitating a microscopic, kinetic-level resolution. In this Letter, we therefore investigate tokamak microwave breakdown by employing \texttt{BREAK}, a three-dimensional Monte Carlo simulation for tokamak breakdown \cite{Yoo2017,Yoo2018}.

A small stray magnetic field\footnote{Here, the stray field denotes the residual magnetic field before plasma formation, distinct from the conventional intrinsic error field \cite{JKPark2007,Boozer2001}.} with a finite amplitude is always present in a plasmaless tokamak and characterizes the open-field magnetic topology. As elucidated later, the magnetic topology plays a key role in the interplay among particle transport, collisions, and nonlinear wave--particle interactions. For transport, the topology governs convection driven by magnetic drifts and vertical diffusion arising from the Brownian motion. For ionization, the non-Maxwellian energy distribution of ionizing particles emerges from the distinct individual histories of energy gain, collisions, and transport that each particle undergoes. Capturing these coupled multiphysics interactions is therefore essential to precisely characterize the ionization source rate and particle loss rate, which together determine the breakdown onset.

To assess the predictive capability of \texttt{BREAK}, we validate the predicted breakdown boundary against controlled, non-inductive KSTAR experiments. Under ITER-relevant conditions at a $\mathrm{D}_2$ prefill pressure \(p_{\mathrm{D}_2}=2\,\mathrm{mPa}\) and a residual poloidal field of \(30\,\mathrm{G}\), an injected EC power \(P_{\mathrm{EC}}=1\,\mathrm{MW}\) gives an ionization source rate above \(6\times10^{3}\,\mathrm{s^{-1}}\). This exceeds the particle loss rate of \(\simeq4.9\times10^{3}\,\mathrm{s^{-1}}\), predicting that microwave breakdown will be feasible in reactor-grade tokamaks.

%
%
%
%
\textit{Physics of tokamak microwave gas breakdown}.---The nonlinear EC energy gain is a fast microscopic process, separated from the slower macroscopic dynamics of transport, ionization, and loss. This separation is implemented in \texttt{BREAK}, which advances particle orbits in the toroidal geometry subject to electron--neutral elastic collisions, ionization, and boundary losses \cite{Yoo2017}. The particle momentum and energy increments are determined from the time-independent resonant Hamiltonian of Ref.~\cite{Farina2018} for each resonant passage through the EC beam. Their evaluation follows the framework of Ref.~\cite{Gwak2025}. We consider second-harmonic extraordinary-mode (X2) heating with near-perpendicular injection from the low-field-side midplane. Further details on the physical assumptions and numerical implementation are given in Sec.~S2 of the Supplemental Material~\cite{SM}. \nocite{Battaglia2019,JayhyunKim2011,YoungOkKim2013}

\begin{figure}[!t]
    \centering  
    \includegraphics[width=\linewidth]{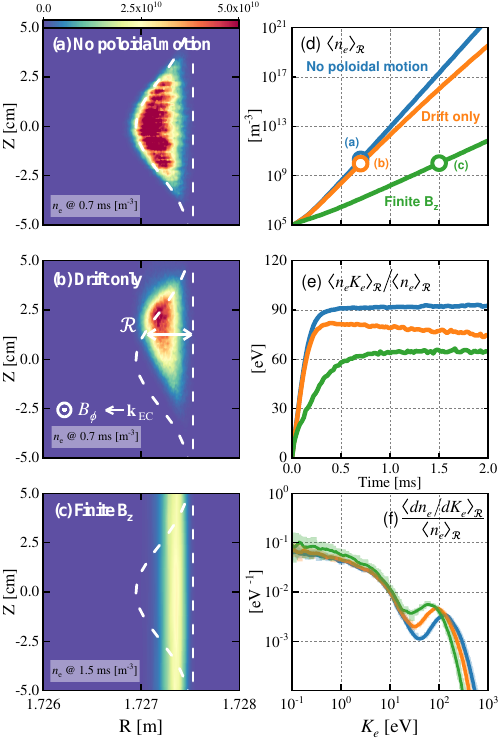}
    \caption{(a)--(c)~Toroidally averaged \(n_e\) for the labeled transport cases at comparable \(\langle n_e\rangle_{\mathcal R}\), with corresponding times marked in (d). White dashed lines mark the cold X2 resonance (vertical) and the boundary for nonlinear EC interaction of cold electrons (curved)~\cite{Farina2019}. (d)~$\langle n_e \rangle_{\mathcal{R}}$ and (e)~$\langle n_e K_e \rangle_{\mathcal{R}}/\langle n_e \rangle_{\mathcal{R}}$ versus time. (f)~Corresponding normalized electron kinetic energy distributions, \(\langle dn_e/dK_e\rangle_{\mathcal{R}}/\langle n_e\rangle_{\mathcal{R}}\). Solid curves show smoothed distributions, and shading shows the local range of the raw data. $\omega/2\pi=140\,\mathrm{GHz}$, $P_{\rm EC}=0.6\,\mathrm{MW}$, beam radius $r=3.3\,\mathrm{cm}$, and toroidal field $B_0=2.4\,\mathrm{T}$ at major radius $R_0=1.8\,\mathrm{m}$, with \(|Z|<1.2\,\mathrm{m}\).
    }
    \label{fig1}
\end{figure}

The effect of particle transport on the ionization avalanche can be isolated by considering controlled cases in which transport effects are selectively included. Figure~\ref{fig1} compares three such cases under KSTAR parameters~\cite{Joung2020FED,Joung2024FED}. With poloidal motion suppressed, the avalanche remains localized in $\mathcal{R}$, the region where nonlinear EC interaction occurs for cold electrons \cite{Farina2019}. $\nabla B$ and curvature drifts alter the spatial distribution. A finite vertical field $B_Z$, used here as a proxy for stray poloidal field, redistributes electrons along open field lines and slows the avalanche. Here and below, \(\langle\cdot\rangle_{\mathcal R}\) denotes a volume average over \(\mathcal R\). The average electron density $\langle n_e \rangle_{\mathcal{R}}$ changes by orders of magnitude across these cases [Fig.~\ref{fig1}(d)], demonstrating the critical role of particle transport in the net avalanche growth. The average electron kinetic energy $\langle n_e K_e \rangle_{\mathcal{R}} / \langle n_e \rangle_{\mathcal{R}}$, by contrast, varies at the tens-of-percent level [Fig.~\ref{fig1}(e)]. Although modest, this variation suggests that macroscopic transport affects the formation of the non-Maxwellian distribution that determines the ionization rate, as demonstrated in Fig.~\ref{fig1}(f). This multiscale feature therefore makes microwave breakdown difficult to characterize by simplified parameters, as in the Townsend avalanche model.

\begin{figure}[!t]
    \centering
    \includegraphics[width=\linewidth]{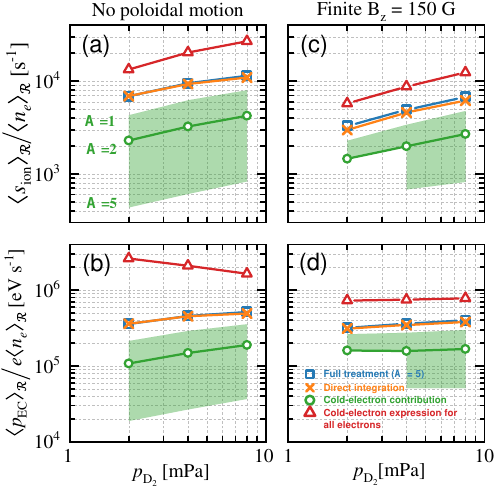}
    \caption{(a),(b)~\(\langle s_{\rm ion}\rangle_{\mathcal R}/\langle n_e\rangle_{\mathcal R}\) and \(\langle p_{\rm EC}\rangle_{\mathcal R}/(e\langle n_e\rangle_{\mathcal R})\) versus \(p_{\mathrm{D}_2}\) with poloidal motion suppressed. (c),(d)~The same quantities at \(B_Z=150\,\mathrm{G}\). The \(\mathcal A=5\) case retaining only the cold-electron contribution does not avalanche at \(B_Z=150\,\mathrm{G}\) and is therefore absent in (c),(d). $\omega/2\pi=170\,\mathrm{GHz}$, $P_{\rm EC}=0.6\,\mathrm{MW}$, $r=14\,\mathrm{cm}$, and $B_0=2.65\,\mathrm{T}$ at $R_0=6.2\,\mathrm{m}$, with \(|Z|<1.6\,\mathrm{m}\).
    }
    \label{fig2}
\end{figure}

We define cold electrons as those satisfying the adiabatic condition and having incident perpendicular kinetic energy of order \(1\,\mathrm{eV}\). Following Ref.~\cite{Johansson2024}, the adiabatic condition corresponds to an incident parallel kinetic energy below \([P_{\rm EC}/(1\,\mathrm{MW})]/(\mathcal A/5)^2\,\mathrm{eV}\), where \(\mathcal{A}\) sets the boundary value for the ratio of the beam flight time to the wave trapping time. Electrons not satisfying these criteria are referred to as hot electrons. The nonlinear EC interaction mechanism by which initially cold electrons gain sufficient energy to drive ionizing collisions during microwave breakdown in tokamaks was originally investigated by Farina \cite{Farina2018}. Building on this understanding, we discover that energy gain of hot electrons accounts for a significant fraction of total energy gain during the ionization avalanche. Figures~\ref{fig2}(a) and \ref{fig2}(b) show $\langle s_{\rm ion}\rangle_{\mathcal{R}}/\langle n_e\rangle_{\mathcal{R}}$ and $\langle p_{\rm EC}\rangle_{\mathcal{R}}/(e\langle n_e\rangle_{\mathcal{R}})$ versus $p_{\mathrm{D}_2}$. Here, $s_{\rm ion}$ is the total ionization rate density from direct and dissociative electron-impact ionization of \(\mathrm{D}_2\), and $p_{\rm EC}$ is the EC power deposition density. Poloidal motion is suppressed in (a) and (b), decoupling the ionization source rate from transport. The magnetic geometry and EC beam parameters are based on ITER plasma initiation conditions, except that perpendicular injection is assumed~\cite{Gribov2018EPS,HyunTaeKim2020,Peter2019,Fanale2021,Moro2020}. Both $\langle p_{\rm EC}\rangle_{\mathcal{R}}/(e\langle n_e\rangle_{\mathcal{R}})$ and the resulting $\langle s_{\rm ion}\rangle_{\mathcal{R}}/\langle n_e\rangle_{\mathcal{R}}$ agree with direct integration of Hamilton's equations of motion in magnitude and in their dependence on prefill pressure when the treatment of Ref.~\cite{Gwak2025} is applied. However, retaining only the contribution from cold electrons underestimates both quantities by a factor of $\sim2$ to more than an order of magnitude depending on $\mathcal{A}$, whereas applying the cold-electron expression to all electrons overestimates them by a factor of \(2\) or more. These differences remain visible at $B_Z=150\,\mathrm{G}$ [Figs.~\ref{fig2}(c),(d)], representative of the vertical-field scale used in the ITER-relevant projection below.

The monotonic increase in $\langle s_{\rm ion}\rangle_{\mathcal{R}}/\langle n_e\rangle_{\mathcal{R}}$ with $p_{\mathrm{D}_2}$ in Fig.~\ref{fig2} implies faster early electron multiplication and hence an earlier onset of the $\mathrm{D}_\alpha$ rise. This prediction is qualitatively consistent with Fig.~11 of Ref.~\cite{Zhang2026}, where the $\mathrm{D}_\alpha$-rise onset time $t_{\rm onset}$ shifts monotonically earlier with increasing $p_{\mathrm{D}_2}$. The comparison is limited to the early breakdown phase and does not extend to the $\mathrm{D}_\alpha$-peak time $t_{\rm peak}$, which can be affected by later $E\times B$ transport and burn-through dynamics not included in the present model.

In Townsend breakdown, the particle loss mechanism is governed by parallel motion of electrons accelerated by an electric field. In microwave breakdown, by contrast, charged particles are transported under the open-field magnetic topology, in which the particles experience \(\nabla B\) and curvature drifts, parallel motion, and collisions without the inductive acceleration. In a realistic tokamak, the open-field topology intrinsically contains the stray magnetic field. Its ratio to the toroidal magnetic field at the cold X2 resonance $B_Z/B_{\rm res}$ is very small, typically less than $1\%$, but the small field line tilt away from the toroidal direction determines a projection angle of parallel motion. Accordingly, the vertically projected particle transport spans three distinct asymptotic regimes as $B_Z/B_{\rm res}$ increases. In this perspective, the role of $B_Z/B_{\rm res}$ in open-field particle transport is analogous to that of the inverse aspect ratio in neoclassical transport with closed flux surfaces: the former controls the vertical displacement, just as the latter determines the banana width characterizing the radial displacement.

\begin{figure}
    \centering
    \includegraphics[width=\linewidth]{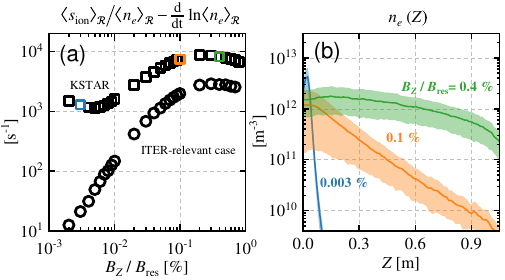}
    \caption{(a)~Particle loss rate versus \(B_Z/B_{\rm res}\) for the KSTAR and ITER-relevant cases. Converged values are used except at low \(B_Z/B_{\rm res}\), where transport remains nonstationary due to magnetic drift motion. The loss rate is then evaluated at \(\langle n_e\rangle_{\mathcal R}=10^{12}\,\mathrm{m}^{-3}\). (b)~\(n_e(Z)\) for the three colored KSTAR cases in (a), at matched \(\langle n_e\rangle_{\mathcal R}=10^{12}\,\mathrm{m}^{-3}\). Solid lines and bands show the \(R\)-averaged profiles and radial variation, respectively.
    }
    \label{fig3}
\end{figure}

Figure~\ref{fig3}(a) compares the particle loss rate $\langle s_{\rm ion}\rangle_{\mathcal{R}}/\langle n_e\rangle_{\mathcal{R}} - d\ln\langle n_e\rangle_{\mathcal{R}}/dt$ as a function of $B_Z/B_{\rm res}$ for the KSTAR and ITER-relevant cases of Figs.~\ref{fig1} and~\ref{fig2}, respectively. In the quasi-toroidal limit $B_Z/B_{\rm res}\ll 1$, $\nabla B$ and curvature drifts predominantly form a unidirectional, \textit{convective} flow of particles. For the KSTAR device, the upward flow drives an outward particle flux from $\mathcal{R}$ [Fig.~\ref{fig1}(b)]. This accounts for the increase in the loss rate when $B_Z/B_{\rm res}$ decreases to sufficiently small values. This drift-dominated regime is not expected to be accessible in the ITER-relevant case over the same \(B_Z/B_{\rm res}\) range because the larger ITER beam radius increases the distance over which magnetic drifts must convect particles out of \(\mathcal{R}\). 

As $B_Z/B_{\rm res}$ increases, the random parallel motion surpasses the drift motion and forms a bidirectional, \textit{diffusive} flow of particles. The parallel spreading of electrons is a Brownian process in the field-line coordinate $s$ with diffusivity $D_s\sim\langle v_\parallel^2\rangle\tau_{\rm coll}/2$, where $\tau_{\rm coll}$ is the characteristic collisional time. When $B_Z/B_{\rm res}$ is moderately small and the parallel motion is purely diffusive, changing $B_Z$ barely influences the nonlinear energy gain (within a few percent) but directly modifies the projection angle. This suggests that the normalized diffusive particle flux along $s$, $|\Gamma_s|/n_e\sim D_s/L_s$, where $L_s^{-1}\equiv|\partial_s\ln n_e|$, develops independently of the magnetic topology. The corresponding vertical diffusivity follows from the magnetic-pitch relation $\Delta Z\simeq(B_Z/B_{\rm res})\Delta s$, giving $D_Z\sim\frac{\left\langle(\Delta Z)^2\right\rangle}{2\tau_{\rm coll}}\sim D_s\left(\frac{B_Z}{B_{\rm res}}\right)^2$, where $\left\langle(\Delta Z)^2\right\rangle$ is the mean square of the vertical displacement of electrons over $\tau_{\rm coll}$. Because $L_Z\simeq L_s(B_Z/B_{\rm res})$ for $L_Z^{-1}\equiv|\partial_Z\ln n_e|$, the diffusive flux scales with $D_Z/L_Z\propto B_Z/B_{\rm res}$, so the loss rate rises approximately linearly in this regime. Indeed, when $B_Z/B_{\rm res} \in (10^{-2}, 10^{-1})\%$, the linear dependence is demonstrated in Fig.~\ref{fig3}(a).

For $B_Z/B_{\rm res} \gtrsim 0.2\%$, the loss rate decreases in both devices, reflecting the boundary effect. When $L_s$ becomes comparable to the connection length $L_c$, the random walk is constrained by the boundaries along $s$, which reduces $D_s/L_s$. The representative green and orange curves in Fig.~\ref{fig3}(b) indicate that $L_Z$ is substantially larger in this constrained-diffusive regime than in the purely diffusive regime.

This transport mechanism is qualitatively consistent with the progressively earlier \(\mathrm{D}_\alpha\) rise observed in DIII-D as the applied vertical field increased in magnitude \cite{Jackson2010FST}. Simulations for DIII-D using a rough estimate $p_{\mathrm{D}_2}\simeq2$--$4\,\mathrm{mPa}$ place the transition to the constrained-diffusive regime near $L_c\simeq660\,\mathrm{m}$, within the range of connection lengths reported for that scan. The weak dependence observed in J-TEXT does not contradict this picture, as \(t_{\rm onset}\) changed little when the vertical-field coil current was varied \cite{Zhang2026}. This weak response can be understood because the ionization source rate likely far exceeded the particle loss rate, making \(t_{\rm onset}\) largely insensitive to changes in transport. These observations are therefore qualitatively consistent with the proposed mechanism, while quantitative validation requires a dedicated controlled experiment.

%
%
%
%
\begin{figure}[t]
    \centering
    \includegraphics[width=\linewidth,trim=0 0 0 0,clip]{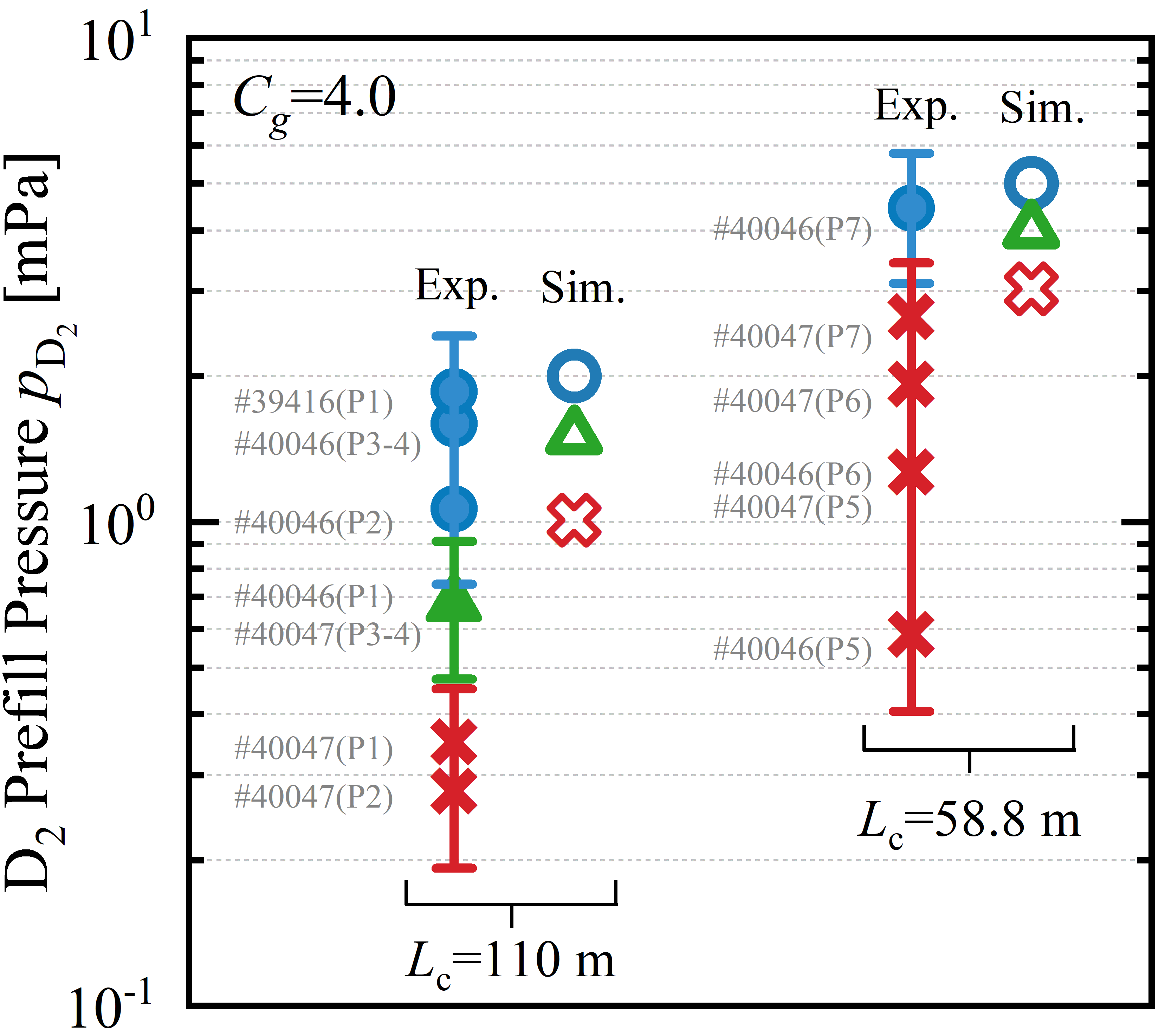}
    \caption{     
            Breakdown results at \(L_c=110\) and \(58.8\,\mathrm{m}\). Circles and crosses denote success and failure. Green triangles denote repeated experimental conditions yielding either failure or breakdown delayed by several hundred milliseconds (filled), and single-pass failure with second-pass success in simulation (open). Vertical bars indicate the ranges corresponding to the \(\pm30\%\) gauge accuracy. Experimental and simulated markers are horizontally offset within each \(L_c\) group for clarity only. Gray labels identify shots and pulses.
            }
    \label{fig5}
\end{figure}
\textit{Experimental validation}.---We performed controlled microwave breakdown experiments in KSTAR at zero loop voltage, using 105-GHz X2 pulses at $0.6\,\mathrm{MW}$ while scanning the $\mathrm{D}_2$ prefill pressure. We used two vertical-field configurations with connection lengths $L_c=110$ and \(58.8\,\mathrm{m}\), corresponding to the diffusive transport regimes. The vertical field \(B_Z\) in Fig.~\ref{fig3} denotes a proxy for the stray fields in a field-null configuration. For the present validation, however, the magnetic configuration is deliberately tailored to reproduce these fields at a controlled magnitude. Breakdown was identified when the raw $\mathrm{D}_\alpha$ signal rose above the noise floor in any toroidal or poloidal channel during the pulse. Further experimental and simulation details are given in Secs.~S1 and S2 of the Supplemental Material~\cite{SM}, respectively.

A simulated case is classified as successful breakdown when the electron population in \(\mathcal R\) exhibits sustained multiplication at a rate \(d\ln\langle n_e\rangle_{\mathcal R}/dt>100\,\mathrm{s^{-1}}\). This threshold translates to a \(10^{10}\) multiplication within approximately \(0.23\,\mathrm{s}\), excluding cases that would break down on an unreasonably long time scale. The EC wave is weakly absorbed on its first pass, and the single-pass treatment may therefore overestimate the prefill pressure required for breakdown. Hence, the resulting estimate remains appropriately conservative for assessing startup scenarios in ITER \cite{Peter2019} and other current \cite{Yang2024,Wakatsuki2024} and future tokamaks \cite{KDEMO_Kim2015,CFETR_Song2022,JADEMO_Tobita2019,EUDEMO_Bachmann2020,DTT_Romanelli2024,COMPASSU_Vondracek2021,IndianDEMO_Deshpande2023}.

Figure~\ref{fig5} compares the experimental classification with the numerical prediction as a function of \(p_{\mathrm{D}_2}\) for the two $L_c$ settings. The in-vessel \(\mathrm{D}_2\) prefill pressure was inferred from the duct-gauge pressure using \texttt{MolFlow+} simulations \cite{Kersevan2009MolflowPlus,Jeong2023} of the KSTAR vacuum geometry, giving \(p_{\mathrm{D}_2}=C_g p_{\mathrm{D}_2,\mathrm{duct}}\) with \(C_g=4\). For both \(L_c\) environments, the simulated and experimental breakdown boundaries agree within a factor of two in prefill pressure. The remaining difference can be attributed to multi-pass effects. Indeed, at the condition marked by the open green triangle, the simulations predict failure under the single-pass treatment but successful breakdown when a second pass is included.

%
%
%
%
\begin{figure}[t]
    \centering
    \includegraphics[width=\linewidth,trim=0 0 0 0,clip]{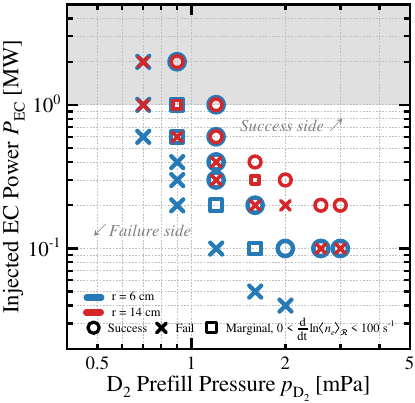}
    \caption{
            Predicted pressure--power scan results for microwave breakdown under ITER-relevant conditions. Scan points adjacent to a change in outcome along either axis are shown for clarity. Gray shading indicates \(P_{\rm EC}>1\,\mathrm{MW}\), the nominal output of a single ITER EC gyrotron~\cite{Oda2019}. Other simulation parameters follow the Fig.~\ref{fig2} baseline.
            }
    \label{fig6}
\end{figure}
\textit{ITER-relevant projection}.---Figure~\ref{fig6} shows the predicted breakdown boundary for the ITER-relevant configuration considered in Fig.~\ref{fig2}. We scan two beam radii, \(r=6\) and \(14\,\mathrm{cm}\), spanning most of the \(5\)--\(15\,\mathrm{cm}\) range specified in the ITER first-plasma EC optical design study~\cite{Moro2020}. In practice, a null configuration has a small stray field across the field null. Hence, an ideal estimate of the connection length \(L_c\) necessitates the full magnetic field configuration. Instead, we simply introduce the proxy vertical field \(B_Z=B_{\rm pol}/f_{\rm null}\simeq170\,\mathrm{G}\), for which the effective connection length \(\widetilde{L}_c=2f_{\rm null}|Z|_{\rm max} B_{\rm res}/B_{\rm pol}\) represents the characteristic loss scale of the ITER field-null startup. Here \(|Z|_{\rm max}=1.6\,\mathrm{m}\) is the vertical half-extent of the simulation domain. The value \(B_{\rm pol}=30\,\mathrm{G}\) denotes the level of the residual poloidal magnetic field that ITER aims to achieve in the breakdown region~\cite{Peter2019}, and \(f_{\rm null}\) is the geometric reduction factor required to compensate for the proxy treatment. Assuming this factor barely varies among tokamaks with field-null configurations, we take \(f_{\rm null}=0.18\) from calibration against the DIII-D breakdown threshold reported in Ref.~\cite{Sinha2022}.

Successful avalanches are obtained near and below \(P_{\rm EC}=1\,\mathrm{MW}\) at 
\(p_{\mathrm{D}_2}\simeq2\,\mathrm{mPa}\), indicating that microwave breakdown under ITER-relevant conditions remains accessible at modest injected EC power. This prediction is less stringent than the extrapolated estimate of \(3.9\,\mathrm{MW}\) based on the \(R_0^2\) scaling in Ref.~\cite{Zhang2026}. The discrepancy appears to arise from their consideration of the post-breakdown phase, in which the power requirement for positive density growth would be more stringent to overcome global heating. The present \texttt{BREAK} simulations only predict successful ``gas'' breakdown below \(1\,\mathrm{MW}\) over the same pressure range. We confirmed that the breakdown remains successful even with a conservative choice of \(f_{\rm null}=0.1\).

%
%
%
%
\textit{Conclusions}.---We investigate microwave gas breakdown in a tokamak by performing self-consistent three-dimensional Monte Carlo simulation with \texttt{BREAK}. The simulation reveals the roles of the stray magnetic field in the particle transport as well as ionizing events. We find drift-dominated, purely diffusive, and constrained-diffusive transport regimes under the open-field magnetic topology, which qualitatively explain the anomalous trend of \(\mathrm{D}_\alpha\) in DIII-D \cite{Jackson2010FST}. For ionization, beyond the intuitive dependence on prefill pressure, capturing the non-Maxwellian energy distribution is essential; this requires resolving both the energy-dependent transport and the energy gain of hot electrons. The understanding of the initial ionization avalanche mechanism will pave the way for elucidating a post-breakdown mechanism in various startup scenarios including a promising trapped particle configuration \cite{JeongwonLee2017,Yoneda2017,Wakatsuki2024,Tsujii2026}, in which \(B_Z\) is deliberately tailored rather than existing as a stray field. The prediction of microwave breakdown onset by \texttt{BREAK} is validated in KSTAR. Relying on its prediction, we conclude that microwave gas breakdown will be feasible under ITER-relevant conditions at a \(\mathrm{D}_2\) prefill pressure near $2\,\mathrm{mPa}$ with $1\,\mathrm{MW}$ of injected EC power.

\begin{acknowledgments}
\textit{Acknowledgments}.---This work was supported by the National R\&D Program through the National Research Foundation of Korea (NRF), funded by the Korean government (Ministry of Science and ICT) (NRF-2021M1A7A4091135), and by the U.S. Department of Energy, Office of Science, Office of Fusion Energy Sciences, under Award No. DE-FG02-95ER54309. This work was also supported by the R\&D Program of ``High Performance Tokamak Plasma Research \& Development'' (code No. EN2501) through the Korea Institute of Fusion Energy (KFE), funded by the Government of the Republic of Korea.
\end{acknowledgments}

\bibliography{jinwoos_bib}

@article{Farina2018,
    author = "Daniela Farina",
    title = "Nonlinear collisionless electron cyclotron interaction in the pre-ionisation stage",
    journal = "Nucl. Fusion",
    volume = 58,
    number = 6,
    pages = 066012,
    year = 2018,
    doi = {10.1088/1741-4326/aabaa7},
    url = {https://doi.org/10.1088/1741-4326/aabaa7},
}

@article{Johansson2024, 
    title={Electron cyclotron resonance during plasma initiation}, 
    volume={90}, 
    DOI={10.1017/S0022377823001423}, 
    number={1}, 
    journal={J. Plasma Phys.}, 
    author={Johansson, C. Albert and Aleynikov, Pavel}, 
    year={2024}, 
    pages={905900103}
}

@article{Stober2011,
    author = "J. Stober and G.L. Jackson and E. Ascasibar and Y.-S. Bae and J. Bucalossi and A. Cappa and T. Casper and M.-H. Cho and Y. Gribov and G. Granucci and others",
    title = "{ECRH}-assisted plasma start-up with toroidally inclined launch: multi-machine comparison and perspectives for {ITER}",
    journal = "Nucl. Fusion",    
    volume = 51,
    number = 8,
    pages = 083031,    
    year = 2011,
    doi = {10.1088/0029-5515/51/8/083031},
    url = {https://doi.org/10.1088/0029-5515/51/8/083031},
}

@article{Sinha2022,
    author = "J. Sinha and P.C. de Vries and M.L. Walker and D.J. Battaglia and F. Turco and A. Hyatt and H.T. Kim and J. Stober and R. Yoneda and Y. Gribov and others",
    title = "Studies of {EC} pre-ionization in {DIII-D} to support development of ITER plasma initiation",
    journal = "Nucl. Fusion",
    volume = 62,
    number = 6,
    pages = 066013,
    year = 2022,    
    doi = {10.1088/1741-4326/ac59ea},
    url = {https://doi.org/10.1088/1741-4326/ac59ea},
}

@article{Battaglia2019,
    doi = {10.1088/1741-4326/ab3bd5},
    url = {https://doi.org/10.1088/1741-4326/ab3bd5},
    year = {2019},
    month = {oct},
    publisher = {IOP Publishing},
    volume = {59},
    number = {12},
    pages = {126016},
    author = {Battaglia, D.J. and Thornton, A.J. and Gerhardt, S.P. and Kirk, A. and Kogan, L. and Menard, J.E.},
    title = {Reduced model for direct induction startup scenario development on {MAST-U} and {NSTX-U}},
    journal = {Nucl. Fusion},
}

@article{JeongwonLee2017,
    author = "Jeongwon Lee and Jayhyun Kim and YoungHwa An and Min-Gu Yoo and Y.S. Hwang and Yong-Su Na",
    title = "Study on {ECH}-assisted start-up using trapped particle configuration in {KSTAR} and application to {ITER}",
    journal = "Nucl. Fusion",
    volume = 57,
    number = 12,
    pages = 126033,
    year = 2017,
    doi = {10.1088/1741-4326/aa8511},
    url = {https://doi.org/10.1088/1741-4326/aa8511},
}

@article{kubo1983,
    title = {Toroidal {P}lasma {C}urrent {S}tartup and {S}ustainment by rf in the {WT}-2 {T}okamak},
    author = {Kubo, S. and Nakamura, M. and Cho, T. and Nakao, S. and Shimozuma, T. and Ando, A. and Ogura, K. and Maekawa, T. and Terumichi, Y. and Tanaka, S.},
    journal = {Phys. Rev. Lett.},
    volume = {50},
    issue = {25},
    pages = {1994--1997},
    numpages = {0},
    year = {1983},
    month = {Jun},
    publisher = {American Physical Society},
    doi = {10.1103/PhysRevLett.50.1994},
    url = {https://link.aps.org/doi/10.1103/PhysRevLett.50.1994}
}

@article{toi1984,
    title = {Startup and {Q}uasistationary {D}rive of {P}lasma {C}urrent by {L}ower {H}ybrid {W}aves in a {T}okamak},
    author = {Toi, K. and Ohkubo, K. and Kawahata, K. and Kawasumi, Y. and Matsuoka, K. and Noda, N. and Ogawa, Y. and Sato, K. and Tanahashi, S. and Tetsuka, T. and others},
    journal = {Phys. Rev. Lett.},
    volume = {52},
    issue = {24},
    pages = {2144--2147},
    numpages = {0},
    year = {1984},
    month = {Jun},
    publisher = {American Physical Society},
    doi = {10.1103/PhysRevLett.52.2144},
    url = {https://link.aps.org/doi/10.1103/PhysRevLett.52.2144}
}

@article{Bae2008,
    author = "Y.S. Bae and J.H. Jeong and S.I. Park and M. Joung and J.H. Kim and S.H. Hahn and S.W. Yoon and H.L. Yang and W.C. Kim and Y.K. Oh and others",
    title = "{ECH} pre-ionization and assisted startup in the fully superconducting {KSTAR} tokamak using second harmonic",
    year = 2008,
    volume = 49,
    number = 2,
    pages = 022001,
    journal = "Nucl. Fusion",
    doi = {10.1088/0029-5515/49/2/022001},
    url = {https://doi.org/10.1088/0029-5515/49/2/022001},
}

@misc{Boozer2026,
    title={Constraints on the magnetic field evolution in tokamak power plants}, 
    author={Allen H Boozer},
    eprint={2507.05456},
    archivePrefix={arXiv},
}

@article{Johansson2026,
    doi = {10.1088/1741-4326/ae2b84},
    url = {https://doi.org/10.1088/1741-4326/ae2b84},
    year = {2026},
    month = {dec},
    publisher = {IOP Publishing},
    volume = {66},
    number = {2},
    pages = {026011},
    author = {Johansson, C. Albert and Aleynikov, Pavel and Cappa, Alvaro and the W7-X Team},
    title = {Model for second harmonic {ECRH} plasma start-up in stellarators},
    journal = {Nucl. Fusion},
}

@article{JKPark2007,
  title = {Control of {A}symmetric {M}agnetic {P}erturbations in {T}okamaks},
  author = {Park, Jong-kyu and Schaffer, Michael J. and Menard, Jonathan E. and Boozer, Allen H.},
  journal = {Phys. Rev. Lett.},
  volume = {99},
  issue = {19},
  pages = {195003},
  numpages = {4},
  year = {2007},
  month = {Nov},
  publisher = {American Physical Society},
  doi = {10.1103/PhysRevLett.99.195003},
  url = {https://link.aps.org/doi/10.1103/PhysRevLett.99.195003}
}

@article{Hwang1996,
  title = {Observation of {N}onclassical {R}adial {C}urrent {D}iffusion in a {F}ully {B}ootstrap {C}urrent {D}riven {T}okamak},
  author = {Hwang, Y. S. and Forest, C. B. and Ono, M.},
  journal = {Phys. Rev. Lett.},
  volume = {77},
  issue = {18},
  pages = {3811--3814},
  numpages = {0},
  year = {1996},
  month = {Oct},
  publisher = {American Physical Society},
  doi = {10.1103/PhysRevLett.77.3811},
  url = {https://link.aps.org/doi/10.1103/PhysRevLett.77.3811}
}

@article{Boozer2001,
  title = {Error {F}ield {A}mplification and {R}otation {D}amping in {T}okamak {P}lasmas},
  author = {Boozer, Allen H.},
  journal = {Phys. Rev. Lett.},
  volume = {86},
  issue = {22},
  pages = {5059--5061},
  numpages = {0},
  year = {2001},
  month = {May},
  publisher = {American Physical Society},
  doi = {10.1103/PhysRevLett.86.5059},
  url = {https://link.aps.org/doi/10.1103/PhysRevLett.86.5059}
}

@article{Fitzpatrick2026,
    doi = {10.1088/1741-4326/ae1308},
    url = {https://doi.org/10.1088/1741-4326/ae1308},
    year = {2025},
    month = {nov},
    publisher = {IOP Publishing},
    volume = {66},
    number = {1},
    pages = {016012},
    author = {Fitzpatrick, Richard},
    title = {A simple model of current ramp-up and ramp-down in tokamaks},
    journal = {Nucl. Fusion},
}

@article{Tsujii2026,
    doi = {10.1088/1741-4326/ae6ab5},
    url = {https://doi.org/10.1088/1741-4326/ae6ab5},
    year = {2026},
    month = {may},
    publisher = {IOP Publishing},
    volume = {66},
    number = {6},
    pages = {066027},
    author = {Tsujii, N. and Ejiri, A. and Shinohara, K. and Peng, Y. and Lin, Y. and Jiang, Z. and Tian, Y. and Adachi, F. and Jiang, Y. and Wang, S. and Yoshida, M. and Takechi, Y.},
    title = {Numerical analysis of electron distribution function under electron cyclotron heating during tokamak start-up},
    journal = {Nucl. Fusion},
}

@article{Peter2019,
    author = "P. C. {De Vries} and Y. Gribov",
    title = "{ITER} breakdown and plasma initiation revisited",
    journal = "Nucl. Fusion",    
    volume = 59,
    number = 9,
    pages = 096043,
    year = 2019,
    doi = {10.1088/1741-4326/ab2ef4},
    url = {https://doi.org/10.1088/1741-4326/ab2ef4},
}

@article{Jackson2007,
    author = "G. L. Jackson and J. S. deGrassie and C. P. Moeller and R. Prater",
    title = "Second harmonic electron cyclotron pre-ionization in the {DIII-D} tokamak",
    journal = "Nucl. Fusion",
    volume = 47,
    number = 4,
    pages = "257-263",
    year = 2007,
    doi = {10.1088/0029-5515/47/4/003},
    url = {https://doi.org/10.1088/0029-5515/47/4/003},
}

@article{Jackson2010FST,
    author = "G. L. Jackson and M. E. Austin and J. S. deGrassie and A. W. Hyatt and J. M. Lohr and T. C. Luce and R. Prater and W. P. West",
    title = "Plasma initiation and start-Up studies in the {DIII-D} tokamak with second-harmonic {EC} assist",
    journal = "Fusion Sci. Technol.",
    volume = 57,
    number = 1,
    pages = 27,
    year = 2010,
    url = "https://doi.org/10.13182/FST10-A9266",
}

@article{Yoo2018,
    author={Min-Gu Yoo and Jeongwon Lee and Young-Gi Kim and Jayhyun Kim and Francesco Maviglia and Adrianus C. C. Sips and Hyun-Tae Kim and Taik Soo Hahm and Yong-Seok Hwang and Hae June Lee and others},
    title="Evidence of a turbulent {ExB} mixing avalanche mechanism of gas breakdown in strongly magnetized systems",
    journal="Nat. Commun.",    
    volume=9,
    number=1,
    pages=3523,
    year=2018,
    doi={10.1038/s41467-018-05839-5},
    url={https://doi.org/10.1038/s41467-018-05839-5}
}

@article{Suvorov1988,
    author = "E. V. Suvorov and M. D. Tokman",
    title = "Generation of accelerated electrons during cyclotron heating of plasmas",
    journal = "Fiz. Plazmy",
    volume = 14,
    number = 8,
    pages = 950,
    year = 1988,
    note = {[Sov. J. Plasma Phys. \textbf{14}, 557 (1988)]}
}

@article{Farina2019,
    author = "D. Farina",
    title = "Electron {C}yclotron collisionless interaction during {EC}-assisted tokamak start-up",
    journal = "EPJ Web Conf.",
    volume = 203,
    pages = 01001,
    year = 2019,
    doi = {10.1051/epjconf/201920301001},
    url = {https://doi.org/10.1051/epjconf/201920301001},
}

@article{Yang2024,
    author = "J. Yang and A.C.C. Sips and P. de Vries and J. Sinha and H.T. Kim and F. Glass and M. Austin and M. van Zeeland and J.L. Herfindal and M. Shafer and others",
    title = "Toroidal injection angle dependence of {EC} assisted plasma initiation at {DIII-D}",
    journal = "Nucl. Fusion",
    year = 2024,
    volume = 64,
    number = 12,
    pages = 126065,
    doi = {10.1088/1741-4326/ad8a6f},
    url = {https://doi.org/10.1088/1741-4326/ad8a6f},
    publisher = {IOP Publishing},
    month = {nov},
}

@article{Yoneda2017,
    author = {Yoneda, R. and Hanada, K. and Nakamura, K. and Idei, H. and Yoshida, N. and Hasegawa, M. and Onchi, T. and Kuroda, K. and Kawasaki, S. and Higashijima, A. and others},
    title = {Effect of magnetic structure on {RF}-induced breakdown in {QUEST}},
    journal = {Phys. of Plasmas},
    volume = {24},
    number = {6},
    pages = {062513},
    year = {2017},
    month = {06},
    issn = {1070-664X},
    doi = {10.1063/1.4985142},
    url = {https://doi.org/10.1063/1.4985142},
}

@article{YeongsunLee2024,
    title = {Binary {N}ature of {C}ollisions {F}acilitates {R}unaway {E}lectron {G}eneration in {W}eakly {I}onized {P}lasmas},
    author = {Lee, Y. and Aleynikov, P. and de Vries, P. C. and Kim, H.-T. and Lee, J. and Hoppe, M. and Park, J.-K. and Choi, G. J. and Gwak, J. and Na, Y.-S.},
    journal = {Phys. Rev. Lett.},
    volume = {133},
    issue = {17},
    pages = {175102},
    numpages = {7},
    year = {2024},
    month = {Oct},
    publisher = {American Physical Society},
    doi = {10.1103/PhysRevLett.133.175102},
    url = {https://link.aps.org/doi/10.1103/PhysRevLett.133.175102}
}

@article{Knoepfel1975,
    title = {High-{E}nergy {R}unaway {E}lectrons in the {O}ak {R}idge {T}okamak},
    author = {Knoepfel, H. and Zweben, S. J.},
    journal = {Phys. Rev. Lett.},
    volume = {35},
    issue = {20},
    pages = {1340--1343},
    numpages = {0},
    year = {1975},
    month = {Nov},
    publisher = {American Physical Society},
    doi = {10.1103/PhysRevLett.35.1340},
    url = {https://link-aps-org-ssl.libproxy.snu.ac.kr/doi/10.1103/PhysRevLett.35.1340}
}

@article{Forest1992,
    title = {Internally {G}enerated {C}urrents in a {S}mall-{A}spect-{R}atio {T}okamak {G}eometry},
    author = {Forest, C. B. and Hwang, Y. S. and Ono, M. and Darrow, D. S.},
    journal = {Phys. Rev. Lett.},
    volume = {68},
    issue = {24},
    pages = {3559--3562},
    numpages = {0},
    year = {1992},
    month = {Jun},
    publisher = {American Physical Society},
    doi = {10.1103/PhysRevLett.68.3559},
    url = {https://link.aps.org/doi/10.1103/PhysRevLett.68.3559}
}

@article{Yoshinaga2006,
  title = {Spontaneous {F}ormation of {C}losed-{F}ield {T}orus {E}quilibrium via {C}urrent {J}ump {O}bserved in an {E}lectron-{C}yclotron-{H}eated {P}lasma},
  author = {Yoshinaga, T. and Uchida, M. and Tanaka, H. and Maekawa, T.},
  journal = {Phys. Rev. Lett.},
  volume = {96},
  issue = {12},
  pages = {125005},
  numpages = {4},
  year = {2006},
  month = {Mar},
  publisher = {American Physical Society},
  doi = {10.1103/PhysRevLett.96.125005},
  url = {https://link.aps.org/doi/10.1103/PhysRevLett.96.125005}
}

@article{Shiraiwa2004,
  title = {Formation of {A}dvanced {T}okamak {P}lasmas without the {U}se of an {O}hmic-{H}eating {S}olenoid},
  author = {Shiraiwa, S. and Ide, S. and Itoh, S. and Mitarai, O. and Naito, O. and Ozeki, T. and Sakamoto, Y. and Suzuki, T. and Takase, Y. and Tanaka, S. and others},
  collaboration = {JT-60 Team},
  journal = {Phys. Rev. Lett.},
  volume = {92},
  issue = {3},
  pages = {035001},
  numpages = {4},
  year = {2004},
  month = {Jan},
  publisher = {American Physical Society},
  doi = {10.1103/PhysRevLett.92.035001},
  url = {https://link.aps.org/doi/10.1103/PhysRevLett.92.035001}
}

@article{Sauter2000,
  title = {Steady-{S}tate {F}ully {N}oninductive {C}urrent {D}riven by {E}lectron {C}yclotron {W}aves in a {M}agnetically {C}onfined {P}lasma},
  author = {Sauter, O. and Henderson, M. A. and Hofmann, F. and Goodman, T. and Alberti, S. and Angioni, C. and Appert, K. and Behn, R. and Blanchard, P. and Bosshard, P. and others},
  journal = {Phys. Rev. Lett.},
  volume = {84},
  issue = {15},
  pages = {3322--3325},
  numpages = {0},
  year = {2000},
  month = {Apr},
  publisher = {American Physical Society},
  doi = {10.1103/PhysRevLett.84.3322},
  url = {https://link-aps-org-ssl.libproxy.snu.ac.kr/doi/10.1103/PhysRevLett.84.3322}
}

@article{Yoo2017,
    author = "Min-Gu Yoo and Jeongwon Lee and Young-Gi Kim and Yong-Su Na",    
    title = "Development of {2D} implicit particle simulation code for ohmic breakdown physics in a tokamak",    
    journal = "Comput. Phys. Commun.",
    volume = 221,
    pages = 143,
    year = 2017,    
    doi = {https://doi.org/10.1016/j.cpc.2017.08.009},
    url = {https://www.sciencedirect.com/science/article/pii/S0010465517302540},
}

@article{YoungOkKim2013,
    title = {Control and operation of the gas injection systems for {KSTAR} tokamak},
    journal = {Fusion Eng. Des.},
    volume = {88},
    number = {6},
    pages = {1132-1136},
    year = {2013},
    note = {Proceedings of the 27th Symposium On Fusion Technology (SOFT-27); Liège, Belgium, September 24-28, 2012},
    issn = {0920-3796},
    doi = {https://doi.org/10.1016/j.fusengdes.2013.01.053},
    url = {https://www.sciencedirect.com/science/article/pii/S092037961300063X},
    author = {Young Ok Kim and Jae In Song and Kwang Pyo Kim and Yong Chu and Kap Rai Park and Hong Tack Kim and Hak Kun Kim and Kun Su Lee and Yang Mo Kim},
}

@article{Moro2020,
    title = {Design of {E}lectron {C}yclotron {R}esonance {H}eating protection components for first plasma operations in {ITER}},
    journal = {Fusion Eng. Des.},
    volume = {154},
    pages = {111547},
    year = {2020},
    issn = {0920-3796},
    doi = {https://doi.org/10.1016/j.fusengdes.2020.111547},
    url = {https://www.sciencedirect.com/science/article/pii/S0920379620300958},
    author = {Alessandro Moro and Alessandro Bruschi and Olivier Darcourt and Francesco Fanale and Daniela Farina and Lorenzo Figini and Franco Gandini and Mark Henderson and Ryan Hunt and Carsten Lechte and others},
}

@article{Oda2019,
    doi = {10.1088/1741-4326/ab22c2},
    url = {https://doi.org/10.1088/1741-4326/ab22c2},
    year = {2019},
    month = {jun},
    publisher = {IOP Publishing},
    volume = {59},
    number = {8},
    pages = {086014},
    author = {Oda, Yasuhisa and Ikeda, Ryosuke and Kajiwara, Ken and Kobayashi, Takayuki and Hayashi, Kazuo and Takahashi, Koji and Moriyama, Shinichi and Sakamoto, Keishi and Eguchi, Taku and Kawakami, Yoshio and others},
    title = {Development of the first {ITER} gyrotron in {QST}},
    journal = {Nucl. Fusion},
}

@article{Joung2020FED,
    title = {Design of {ECH} launcher for {KSTAR} advanced {T}okamak operation},
    journal = {Fusion Eng. Des.},
    volume = {151},
    pages = {111395},
    year = {2020},
    issn = {0920-3796},
    doi = {https://doi.org/10.1016/j.fusengdes.2019.111395},
    url = {https://www.sciencedirect.com/science/article/pii/S0920379619308919},
    author = {Mi Joung and Minho Woo and Jongwon Han and Sonjong Wang and Sunggug Kim and Sanghee Hahn and Dongjea Lee and Jonggu Kwak and Robert Ellis},
}

@article{Joung2024FED,
    title = {Design and operation results of {KSTAR} {ECH} system},
    journal = {Fusion Eng. Des.},
    volume = {203},
    pages = {114461},
    year = {2024},
    issn = {0920-3796},
    doi = {https://doi.org/10.1016/j.fusengdes.2024.114461},
    url = {https://www.sciencedirect.com/science/article/pii/S0920379624003144},
    author = {Mi Joung and Sonjong Wang and Sunggug Kim and Jongwon Han and Inhyuk Rhee and Jonggu Kwak},
}

@article{Fanale2021,
    title = {Design validation of in-vessel mirrors and beam dump for first plasma operations in {ITER}},
    journal = {Fusion Eng. Des.},
    volume = {172},
    pages = {112717},
    year = {2021},
    issn = {0920-3796},
    doi = {https://doi.org/10.1016/j.fusengdes.2021.112717},
    url = {https://www.sciencedirect.com/science/article/pii/S0920379621004932},
    author = {F. Fanale and A. Bruschi and O. Darcourt and D. Farina and L. Figini and F. Gandini and M. Henderson and R. Hunt and C. Lechte and A. Moro and others},
}

@article{Zhang2026,
    doi = {10.1088/1741-4326/ae4fde},
    url = {https://doi.org/10.1088/1741-4326/ae4fde},
    year = {2026},
    month = {apr},
    publisher = {IOP Publishing},
    volume = {66},
    number = {5},
    pages = {056010},
    author = {Zhang, Junli and de Vries, Peter C. and Yang, James and Yang, Zhoujun and Cheng, Zhifeng and Xia, Donghui and Xu, Xin and Yang, Qinhu and Gao, Li and Wang, Nengchao and others},
    title = {Experimental study of {ECH} pre-ionization on {J-TEXT}},
    journal = {Nuclear Fusion},
}

@article{Wakatsuki2024,
    author = "T. Wakatsuki and H. Urano and M. Yoshida and N. Tsujii and S. Inoue and S. Kojima and T. Nakano and M. Fukumoto and Y. Ohtani and R. Sano and others",
    title = "Achievement of the first tokamak plasma with low inductive electric field in {JT-60SA}",
    journal = "Nucl. Fusion",
    year = 2024,
    volume = 64,
    number = 10,
    pages = 104003,
    doi = {10.1088/1741-4326/ad75a7},
    url = {https://doi.org/10.1088/1741-4326/ad75a7},
}

@article{JayhyunKim2011,
    author = "Jayhyun Kim and S.W. Yoon and Y.M. Jeon and J.A. Leuer and N.W. Eidietis and D. Mueller and S. Park and Y.U. Nam and J. Chung and K.D. Lee and others",
    title = "Stable plasma start-up in the {KSTAR} device under various discharge conditions",
    journal = "Nucl. Fusion",
    volume = 51,
    number = 8,    
    year = 2011,    
    pages = 083034,
    doi = {10.1088/0029-5515/51/8/083034},
    url = {https://doi.org/10.1088/0029-5515/51/8/083034},
}

@article{HyunTaeKim2020,
    author = {Kim, Hyun-Tae and Mineev, A. and Ricci, D. and Lee, Jeong-Won and Na, Yong-Su and ITPA-IOS members and JET contributors},    
    title = {Benchmarking of codes for plasma burn-through in tokamaks},
    journal = {Nucl. Fusion},
    doi = {10.1088/1741-4326/abb95c},
    url = {https://doi.org/10.1088/1741-4326/abb95c},
    year = {2020},
    month = {nov},
    publisher = {IOP Publishing},
    volume = {60},
    number = {12},
    pages = {126049},    
}

@inproceedings{Gribov2018EPS,
    author    = {Gribov, Y. and Kavin, A. A. and Lukash, V. E. and Lobanov, K. M. and Mineev, A. B. and Dubrov, M. L. and Khayrutdinov, R. R. and Snipes, J. A. and de Vries, P. C.},
    title     = {Progress in simulation of {ITER} First Plasma operation},
    booktitle = {Proceedings of the 45th {EPS} Conference on Plasma Physics},
    year      = {2018},
    address   = {Prague, Czech Republic},
    month     = jul,
    pages     = {P1.1075},
    url       = {https://lac913.epfl.ch/epsppd3/2018/pdf/P1.1075.pdf}
}

@article{Gwak2025,
    author = "Gwak, Jinwoo and Yoo, Min-Gu and Kim, Hyun-Tae and Lee, Yeongsun and Lee, Jeongwon and Na, Yong-Su",
    title = "Modelling of electron cyclotron energy gain in the tokamak pre-ionization phase",
    journal = "Nucl. Fusion",
    volume = 65,
    number = 5,
    pages = "056038",
    year = 2025,  
    doi = {10.1088/1741-4326/adc9c4},
    url = {https://doi.org/10.1088/1741-4326/adc9c4},
}

@article{Jeong2023,
    title = {Development of {P}enning ion gauge for in-situ measurement of neutral pressure in {VEST}},
    journal = {Fusion Eng. Des.},
    volume = {197},
    pages = {114034},
    year = {2023},
    issn = {0920-3796},
    doi = {https://doi.org/10.1016/j.fusengdes.2023.114034},
    url = {https://www.sciencedirect.com/science/article/pii/S0920379623006142},
    author = {Won Ik Jeong and Yun Ho Jung and June Young Kim and Ki Hyun Lee and Jong Yoon Park and Y．S Hwang},
}

@article{Kersevan2009MolflowPlus,
  author  = {Kersevan, Roberto and Pons, {J.-L.}},
  title   = {Introduction to {MOLFLOW+}: New graphical processing unit-based {Monte Carlo} code for simulating molecular flows and for calculating angular coefficients in the compute unified device architecture environment},
  journal = {Journal of Vacuum Science \& Technology A},
  volume  = {27},
  number  = {4},
  pages   = {1017--1023},
  year    = {2009},
  doi     = {10.1116/1.3153280}
}

@article{KDEMO_Kim2015,
    doi = {10.1088/0029-5515/55/5/053027},
    url = {https://doi.org/10.1088/0029-5515/55/5/053027},
    year = {2015},
    month = {apr},
    publisher = {IOP Publishing},
    volume = {55},
    number = {5},
    pages = {053027},
    author = {Kim, K. and Im, K. and Kim, H.C. and Oh, S. and Park, J.S. and Kwon, S. and Lee, Y.S. and Yeom, J.H. and Lee, C. and Lee, G-S. and others},
    title = {Design concept of {K-DEMO} for near-term implementation},
    journal = {Nucl. Fusion},
}

@article{CFETR_Song2022,
    title = {Engineering design of the {CFETR} machine},
    journal = {Fusion Eng. Des.},
    volume = {183},
    pages = {113247},
    year = {2022},
    issn = {0920-3796},
    doi = {https://doi.org/10.1016/j.fusengdes.2022.113247},
    url = {https://www.sciencedirect.com/science/article/pii/S0920379622002411},
    author = {Yuntao Song and Jiangang Li and Yuanxi Wan and Yong Liu and Xiaolin Wang and Baonian Wan and Peng Fu and Peide Weng and Songtao Wu and Xuru Duan and others},
}

@article{COMPASSU_Vondracek2021,
    title = {Preliminary design of the {COMPASS} upgrade tokamak},
    journal = {Fusion Eng. Des.},
    volume = {169},
    pages = {112490},
    year = {2021},
    issn = {0920-3796},
    doi = {https://doi.org/10.1016/j.fusengdes.2021.112490},
    url = {https://www.sciencedirect.com/science/article/pii/S0920379621002660},
    author = {P. Vondracek and R. Panek and M. Hron and J. Havlicek and V. Weinzettl and T. Todd and D. Tskhakaya and G. Cunningham and P. Hacek and J. Hromadka and others},
}

@article{DTT_Romanelli2024,
    doi = {10.1088/1741-4326/ad5740},
    url = {https://doi.org/10.1088/1741-4326/ad5740},
    year = {2024},
    month = {sep},
    publisher = {IOP Publishing},
    volume = {64},
    number = {11},
    pages = {112015},
    author = {Romanelli, Francesco and on behalf of DTT Contributors and Abate, D. and Acampora, E. and Agguiaro, D. and Agnello, R. and Agostinetti, P. and Agostini, M. and Aimetta, A. and Albanese, R. and others},
    title = {Divertor {T}okamak {T}est facility project: status of design and implementation},
    journal = {Nucl. Fusion},
}

@article{IndianDEMO_Deshpande2023,
    doi = {10.1088/1741-4326/ad0797},
    url = {https://doi.org/10.1088/1741-4326/ad0797},
    year = {2023},
    month = {nov},
    publisher = {IOP Publishing},
    volume = {63},
    number = {12},
    pages = {126060},
    author = {Deshpande, S.P. and Maya, P.N.},
    title = {A staged approach to {I}ndian {DEMO}},
    journal = {Nucl. Fusion},
}

@article{EUDEMO_Bachmann2020,
    title = {Key design integration issues addressed in the {EU DEMO} pre-concept design phase},
    journal = {Fusion Eng. Des.},
    volume = {156},
    pages = {111595},
    year = {2020},
    issn = {0920-3796},
    doi = {https://doi.org/10.1016/j.fusengdes.2020.111595},
    url = {https://www.sciencedirect.com/science/article/pii/S0920379620301435},
    author = {C. Bachmann and S. Ciattaglia and F. Cismondi and G. Federici and T. Franke and C. Gliss and T. Härtl and G. Keech and R. Kembleton and F. Maviglia and others},
}

@article{JADEMO_Tobita2019,
    author = {Kenji Tobita and Ryoji Hiwatari and Yoshiteru Sakamoto and Youji Someya and Nobuyuki Asakura and Hiroyasu Utoh and Yuya Miyoshi and Shinsuke Tokunaga and Yuki Homma and Satoshi Kakudate and others},
    title = {Japan’s {E}fforts to {D}evelop the {C}oncept of {JA DEMO} {D}uring the {P}ast {D}ecade},
    journal = {Fusion Sci. Technol.},
    volume = {75},
    number = {5},
    pages = {372--383},
    year = {2019},
    publisher = {Taylor \& Francis},
    doi = {10.1080/15361055.2019.1600931},
    URL = {https://doi.org/10.1080/15361055.2019.1600931},
    eprint = {https://doi.org/10.1080/15361055.2019.1600931}
}

@misc{SM,
    note = {See Supplemental Material at [URL will be inserted by publisher] for details of the KSTAR experiment, simulation model, and numerical implementation, which includes Refs.~\cite{Battaglia2019,JayhyunKim2011,YoungOkKim2013}.},
}

\end{document}


\title{Supplemental Material for ``Mechanism of Ionization Avalanche in Tokamak Microwave Gas Breakdown''}

\author{Jinwoo Gwak\orcidlink{0000-0003-4767-3328}}
\affiliation{Department of Nuclear Engineering, Seoul National University, Seoul, Republic of Korea}

\author{Yeongsun Lee\orcidlink{0000-0003-4474-416X}}
\affiliation{Department of Nuclear Engineering, Seoul National University, Seoul, Republic of Korea}
\affiliation{Nuclear Research Institute for Future Technology and Policy, Seoul National University, Seoul, Republic of Korea}

\author{Jeongwon Lee\orcidlink{0000-0002-2353-2603}}
\affiliation{Korea Institute of Fusion Energy, Daejeon, Republic of Korea}

\author{Won Ik Jeong\orcidlink{0000-0001-8475-2610}}
\affiliation{Korea Institute of Fusion Energy, Daejeon, Republic of Korea}

\author{Hyun-Tae Kim\orcidlink{0009-0008-2549-5624}}
\affiliation{United Kingdom Atomic Energy Authority, Culham Campus, Abingdon, Oxfordshire OX14 3DB, United Kingdom}

\author{Yong-Seok Hwang}
\affiliation{Department of Nuclear Engineering, Seoul National University, Seoul, Republic of Korea}

\author{Min-Gu Yoo\orcidlink{0000-0002-9244-7066}}
\email[Contact author: ]{yoom@fusion.gat.com}
\affiliation{General Atomics, San Diego, CA 92186-5608, USA}

\author{Yong-Su Na\orcidlink{0000-0001-7270-3846}}
\email[Contact author: ]{ysna@snu.ac.kr}
\affiliation{Department of Nuclear Engineering, Seoul National University, Seoul, Republic of Korea}

\date{September 17, 2026}

\maketitle

%
%
%
%
\section{KSTAR experimental setup}
\begin{figure}[!t]
    \centering
    \includegraphics[width=\linewidth,trim=0 0 0 0,clip]{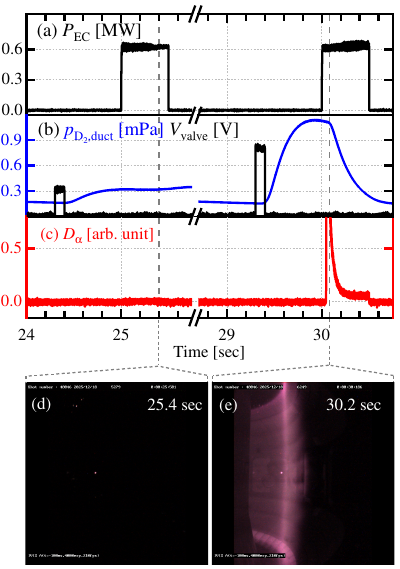}
    \caption{     
            Representative time window from KSTAR shot \#40046. (a)~Injected EC power $P_{\mathrm{EC}}$. (b)~$\mathrm{D}_2$ pressure measured in the vacuum duct in front of the gas pump (blue) and gas-puff valve voltage $V_{\mathrm{valve}}$ (black). (c)~Raw $\mathrm{D}_{\alpha}$ signal from a representative channel. The broken time axis shows two successive EC pulses within the same discharge. (d),(e)~Visible-light camera frames at the times marked by dashed lines in (a)--(c), showing no plasma formation and successful breakdown, respectively.             
            }
    \label{figS1}
\end{figure}

Figure~\ref{figS1} shows a representative time window of a KSTAR discharge in which successive electron cyclotron (EC) pulses, each preceded by a programmed $\mathrm{D}_2$ puff, were applied to test microwave breakdown under zero loop voltage. The toroidal magnetic field was fixed at $B_0=1.8\,\mathrm{T}$ at $R_0=1.8\,\mathrm{m}$. The externally imposed vacuum poloidal field was also held fixed during each EC pulse. EC power was supplied by a single 105-GHz gyrotron and launched in X-mode from the low-field side. The launched power was $0.6\,\mathrm{MW}$ for $0.5\,\mathrm{s}$ per pulse, with $r=5\,\mathrm{cm}$, as estimated by a dedicated ray-tracing calculation. The toroidal launch angle was $\approx5^\circ$, corresponding to near-normal incidence chosen to protect in-vessel components. The poloidal steering was chosen so that the beam intersected the X2 resonance layer near the midplane. With this fixed launch geometry, prefill-pressure scans were performed at two designed-$L_c$ environments. To control $L_c$ while minimizing possible \(E\times B\)-transport effects~\cite{Yoo2018,Battaglia2019}, we used two vertical-field configurations. The corresponding connection lengths, calculated with the \texttt{FIST99} code~\cite{JayhyunKim2011}, were $L_c=110$ and \(58.8\,\mathrm{m}\). The prefill pressure was scanned by varying the gas-puff valve voltage [Fig.~\ref{figS1}(b)]~\cite{YoungOkKim2013}. Breakdown was identified when the raw $\mathrm{D}_\alpha$ signal rose above the noise floor in any toroidal or poloidal channel during the pulse [Fig.~\ref{figS1}(c)]. Representative failed and successful pulses are shown in Figs.~\ref{figS1}(c)--(e).
%
%
%
%
\section{Simulation model and numerical implementation}
\subsection{General physical assumptions}
The electron density remains low throughout the initial phase of microwave breakdown, despite the ionization avalanche, and so does the ionized fraction. Under this condition, we assume that self-generated electrostatic fields are negligible. This approximation remains self-consistent, as such fields become important in KSTAR breakdown simulations only after \(n_e\) approaches \(\mathcal{O}(10^{12})\,\mathrm{m^{-3}}\)~\cite{Yoo2018}. The small ionized fraction also makes neutral depletion negligible. Coulomb collisions are likewise omitted. The elastic and inelastic electron--neutral collisions are modeled as in the original \texttt{BREAK} description of Ref.~\cite{Yoo2017}.

The initial plasma density and temperature are assumed to be \(10^{5}\)--\(10^{6}\,\mathrm{m^{-3}}\) and the room-temperature value of \(0.026\,\mathrm{eV}\), respectively, with the seed electrons distributed uniformly within the device. The physical mechanisms and general trends identified in this work are insensitive to the magnitude of the initial plasma density. The ionization avalanche simulations evolve toward a quasi-stationary state in which the total electron density grows exponentially while its normalized distribution approaches a fixed profile. The rate quantities in Figs.~2 and~3 of the main text are evaluated in this state, except at the low \(B_Z/B_{\rm res}\) points of Fig.~3, as noted therein.

\subsection{Nonlinear EC wave--particle interaction}
The absorbed EC power remains a negligible fraction of the launched power, even though the energy transferred to a resonant electron exceeds the ionization potential. The plasmaless condition \(\omega_{pe}\ll\omega\), where \(\omega_{pe}\) is the electron plasma frequency, fixes the wave amplitude and polarization. The injected EC beam is assumed to be a toroidally localized Gaussian beam. The distance traversed by an electron through the EC beam is much shorter than the electron--neutral ionization mean free path over the range of neutral gas pressures of interest. This scale separation allows the localized nonlinear EC interaction to be treated separately from the slower dynamics of transport, ionization, and loss.

The electron momentum and energy increments at each resonant passage follow from the time-independent resonant Hamiltonian in the wave frame of Ref.~\cite{Farina2018},
\begin{flalign} 
    H(z,\theta;\bar{P}_z,I)=&H_0(\bar{P}_{z},I)+H_1(z,\theta;\bar{P}_z,I),\\
    H_0=&\Gamma-\nu_{n}I,\quad H_1=\epsilon(z)\frac{\Theta_n}{\Gamma}\cos n\theta. \label{eq:Ham2}
\end{flalign}
Here, \(\epsilon(z)=|e|E_{1,M}/(m_e\omega c)\exp[-(cz/\Omega_{ce})^2/r_\parallel^2]\), \(\Gamma=[1+2I+(\bar{P}_z+\nu_n N_\parallel I)^2]^{1/2}\), \(b=\nu N_\perp\sqrt{2I}\), and \(\nu=\omega/\Omega_{ce}=n\nu_n\), with \(n\) the cyclotron harmonic number. The parallel and perpendicular refractive-index components are \(N_{\parallel,\perp}=k_{\parallel,\perp}c/\omega\), and \(\Theta_n\) is the wave polarization factor weighted by the Bessel functions \(J_{n-1}(b)\), \(J_n(b)\), and \(J_{n+1}(b)\). The variables \(z\), \(\theta\), \(I\), and \(\bar{P}_z\) denote the local coordinate measured from the beam center along the magnetic field, the gyrophase in the rotating wave frame, the relativistic perpendicular action, and the transformed parallel canonical momentum, respectively; \(z\), \(I\), and \(\bar{P}_z\) are normalized by \(c\,\Omega_{ce}^{-1}\), \(m_e c^2\,\Omega_{ce}^{-1}\), and \(m_e c\). The quantity \(\Omega_{ce}\) is the nonrelativistic electron cyclotron frequency evaluated at the guiding-center position. For a Gaussian beam, \(E_{1,M}=2[Z_0P_{\rm EC}/(\pi r^2)]^{1/2}\) and \(r_\parallel=r(1-N_\parallel^2)^{-1/2}\), where \(E_{1,M}\) is the peak wave electric field, \(Z_0\) is the vacuum impedance, \(r\) is the Gaussian beam radius at resonance, and \(r_\parallel\) is its projection along the local \(z\) direction, which reduces to \(r\) for the near-perpendicular injection considered here. 

Following Ref.~\cite{Gwak2025}, the increments are obtained complementarily from analytic or semi-analytic expressions in their asymptotic ranges and from direct integration of Hamilton's equations elsewhere. The incident perpendicular kinetic energy cutoff for cold electrons is set to \(3\,\mathrm{eV}\).

The Gaussian beam is further approximated to be circularly symmetric, with beam radius \(r\). In addition to the variation along the local field-aligned coordinate \(z\) contained in \(\epsilon(z)\), the wave amplitude has the same Gaussian profile in the vertical direction, \(E(Z)\propto \exp(-Z^2/r^2)\).

\begin{figure}[!t]
    \centering  
    \includegraphics[width=\linewidth]{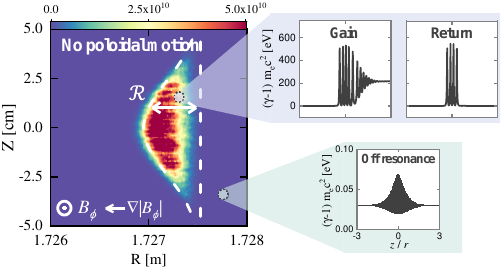}
    \caption{Representative EC wave--particle interactions in the configuration of Fig.~1(a) of the main text. Insets show the kinetic energy evolution along \(z/r\) for initially cold electrons. Passages within \(\mathcal R\) can result in either net energy gain or return close to the incident energy, whereas an off-resonant passage outside \(\mathcal R\) produces negligible energy variation.
    }
    \label{figS2}
\end{figure}

Figure~\ref{figS2} illustrates representative wave--particle interactions for initially cold electrons, showing their kinetic energy evolution along \(z\) during passage through the Gaussian beam. Within \(\mathcal R\), nonlinear resonant interaction can result either in a finite net energy gain or in a return close to the incident energy, whereas outside \(\mathcal R\) the passage is off resonant and produces negligible energy variation.

\subsection{KSTAR validation simulations}
The simulations reproduced the experimental configuration described above on a domain of vertical extent \(|Z|<1.2\,\mathrm{m}\), with a uniform, static \(B_Z\) matched to the field at the X2 resonance for each vertical-field configuration. The wave incidence was taken to be perpendicular because the experimental launch satisfies \(|N_\parallel|\lesssim0.1\) at the resonance, for which the correction to the cold-electron energy gain is below \(6\%\) and is therefore neglected.

For both the KSTAR validation cases and the ITER-like projection, the minimum achievable \(B_Z/B_{\rm res}\) lies above the drift-dominated regime. Residual poloidal magnetic fields in practical startup scenarios are expected to be of order \(0.1\%\) of \(B_{\rm res}\)~\cite{Peter2019}, well within this range. Transport characteristics relevant to such scenarios can therefore be inferred from this controlled KSTAR experiment. The breakdown criterion used in this regime is given in the main text.

\bibliography{jinwoos_bib}